# Three Dimensional Photothermal Deflection of Solids Using Modulated CW Lasers : Theoretical Development

**M. Soltanolkotabi and M. H. Naderi**
**Physics Department , Faculty of Sciences , University of Isfahan, Isfahan, Iran**

**Abstract**

In this paper, a detailed theoretical treatment of the three dimensional photothermal deflection ,under modulated cw excitation , is presented for a three layer system ( backing-solid sample-fluid). By using a technique based on Green's function and integral transformations we find the explicit expressions for laser induced temperature distribution function and the photothermal deflection of the probe beam. Numerical analysis of those expressions for certain solid samples leads to some interesting results.

# I. Introduction

Photothermal techniques evolved from the development of photoacoustic spectroscopy [1] in the 1970s. They now encompass a wide range of techniques and phenomena based upon the conversion of absorbed optical energy into heat . When an energy source is focused on the surface of a sample , part or all of the incident energy is absorbed by the sample and a localized heat flow is produced in the medium following a series of nonradiative deexcitation transitions . Such processes are the origins of the photothermal effects and techniques . If the energy source is modulated, a periodic heat flow is produced at the sample . The resulting periodic heat flow in the material is a diffusive process that produces a periodic temperature distribution called a thermal wave [2].

Several mechanisms are available for detecting , directly or indirectly , thermal waves . These includes gas-microphone photoacoustic detection of heat flow from the sample to the surrounding gas [1] ; photothermal measurements of infrared radiation emitted from the heated sample surface [3] ; optical beam deflection of a laser beam traversing the periodically heated gaseous or liquids layer just above the sample surface [4,5] ; laser detection of the local thermoelastic deformations of the surface [6,7] ; and interferometric detection of the thermoelastic displacement of the sample surface [7,8] . In particular, the two last schemes of thermal wave detection which form the basis of  photothermal deformation spectroscopy (PTDS) [10,11] are becoming the most widely exploited , the principal reason being that they offer a valuable mean for measuring optical and thermal parameters of materials , such as optical absorption coefficient [12] and thermal diffusivity [13].

The photothermal deformation technique is simple and straightforward. A laser beam (pump beam) of wavelength within the absorption range of the sample is incident on the sample and it is  absorbed . The sample gets heated and this heating leads , through thermoelastic coupling , to an expansion of the interaction volume which in turn causes the deformation of the sample surface. The resulting thermoelastic deformation of the surface is detected by the deflection of a  second , weaker laser beam (probe beam).

Ameri and his colleagues were among those who have used first , both the laser interferometric and laser deflection techniques for spectroscopic studies on amorphous silicon [6].Their method restricted to low to moderate modulation frequencies. Opsal *et al* have used thermal wave detection for thin film thickness measurements using laser beam deflection technique[8]. They have obtained temperature distribution function for what is called 1-D temperature distribution function. In their analysis they assumed both probe and pump beams incident normal to the sample surface. Miranda obtained sample temperature distribution by neglecting transient as well as the dc components of the temperature distribution function [9].  On the other hand , the theory of photothermal displacement under pulsed laser excitation in the quasistatic approximation has been given by Li [14] . Zhang and collaborators have investigated the more general case of dynamic thermoelastic response under short laser pulse excitation [15]. Moreover, Cheng and Zhang have considered the effect of the diffusion of photo-generated carriers in semiconductors on the photothermal signal [16].

In this paper , we present a detailed theoretical analysis of the deflection process in three dimensions,for a three layer system consisting of a transparent fluid ,an optically absorbing solid sample, and a backing material. It is assumed that the system is irradiated by a modulated cw laser beam. The theoretical treatment of photothermal deflection can be devided into two parts. In Sec II.A we will find the 3D laser-induced temperature distribution within the three region of the system, due to the absorption of the pump beam. Our mathematical approach is based on Green's function and integral transformations. In Sec II.B the temperature distribution within the fluid derived in Sec II.A will be used to calculate the photothermal deflection of the probe beam directed through the fluid. In Sec III numerical results are presented for typical solid samples, and finally the conclusions are drawn in Sec IV.

**II. Theory of CW Photothermal Deflection**

Let us consider the geometry as shown in Fig.1. The solid sample is assumed to be deposited on a backing and is in contact with a fluid. $l_f$, $l$, and $l_b$ are the thicknesses of the fluid, sample, and the backing, respectively. The fluid can be air or another medium. It is assumed that the solid sample is the only absorbing medium; the fluid and the backing are transparent. For simplicity, we also assume that all three regions extend to infinity in radial directions. A modulated cw cylindrical laser beam irradiates perpendicularly the surface of sample. The first task is to derive the temperature distribution in the fluid due to the heating of the sample surface.

**1st. Temperature Distribution**

The complex amplitude of temperature $\Phi$ is given by a set of 3D heat diffusion equation for the three regions:

$$\frac{\partial^2 \Phi_f}{\partial r^2} + \frac{1}{r}\frac{\partial \Phi_f}{\partial r} + \frac{\partial^2 \Phi_f}{\partial z^2} = \frac{1}{D_f}\frac{\partial \Phi_f}{\partial t} \qquad 0 \leq z \leq l_f \qquad (1)$$

$$\frac{\partial^2 \Phi_s}{\partial r^2} + \frac{1}{r}\frac{\partial \Phi_s}{\partial r} + \frac{\partial^2 \Phi_s}{\partial z^2} = \frac{1}{D_s}\frac{\partial \Phi_s}{\partial t} - A(r)e^{\alpha z}\left(1 + e^{i\omega t}\right) \qquad -l \leq z \leq 0 \qquad (2)$$

$$\frac{\partial^2 \Phi_b}{\partial r^2} + \frac{1}{r}\frac{\partial \Phi_b}{\partial r} + \frac{\partial^2 \Phi_b}{\partial z^2} = \frac{1}{D_b}\frac{\partial \Phi_b}{\partial t} \qquad -l - l_b \leq z \leq -l \qquad (3)$$

In the above equations, $D_f$ $D_s$ and $D_b$ are the thermal diffusivities of the fluid, solid sample and the backing respectively, $\alpha$ is the optical absorption coefficient of the sample, $\omega$ is the frequency of modulation, and $A(r)$ is related to the laser intensity distribution function and is given by

$$A(r) = \frac{\alpha P \eta}{k_s \pi a^2} \exp(-2r^2/a^2). \qquad (4)$$

Here, P is the pump power, $\eta$ is radiation-to-heat conversion efficiency, $k_s$ is the thermal conductivity of the sample and a is the $1/e^2$ radius of the Gaussian pump beam.

Physical constraints on the system manifest as boundary conditions. First, the temperature must be continuous across the region boundaries,

$$\Phi_s\big|_{z=-l} = \Phi_b\big|_{z=-l} \quad , \quad \Phi_s\big|_{z=0} = \Phi_f\big|_{z=0}. \qquad (5)$$

Furthermore, it is assumed that the temperature vanishes far from the sample, i.e;

$$\Phi_b\big|_{z=-\infty} = \Phi_f\big|_{z=+\infty} = 0 \qquad (6)$$

Finally, the heat continuity equation, which states that the heat flux out of one region must equal that into the adjoining region , must be obeyed;

$$k_s \frac{\partial \Phi_s}{\partial z}\bigg|_{z=0} = k_f \frac{\partial \Phi_f}{\partial z}\bigg|_{z=0} \;, \quad k_s \frac{\partial \Phi_s}{\partial z}\bigg|_{z=-l} = k_b \frac{\partial \Phi_b}{\partial z}\bigg|_{z=-l} \;. \tag{7}$$

We assume the pump beam intensity to be sinusoidally modulated for convenience in detection. Therefore in Eq.(2) the source term should be of the form $A(r)\, e^{\alpha z}(1+e^{i\omega t})$. In the steady state , the solutions of Eqs.(1)-(3) contain both static and periodic terms . We will determine the periodic solution only since the signal observed in the laboratory is related to periodic term only, if a phase-sensitive detection is done.

To solve Eqs.(1)-(3) with boundary conditions given by Eqs.(5)-(7) we first apply Hankel transformation. The diffusion equations then become

$$-\lambda^2 \Psi_f + \frac{\partial^2 \Psi_f}{\partial z^2} = \frac{1}{D_f}\frac{\partial \Psi_f}{\partial t} \qquad\qquad 0 \leq z \leq l_f \tag{8}$$

$$-\lambda^2 \Psi_s + \frac{\partial^2 \Psi_s}{\partial z^2} = \frac{1}{D_s}\frac{\partial \Psi_s}{\partial t} - A(\lambda)e^{\alpha z}e^{i\omega t} \qquad -l \leq z \leq 0 \tag{9}$$

$$-\lambda^2 \Psi_b + \frac{\partial^2 \Psi_b}{\partial z^2} = \frac{1}{D_b}\frac{\partial \Psi_b}{\partial t} \qquad\qquad -l-l_b \leq z \leq -l \tag{10}$$

where $\Psi(\lambda,z,t)$ and $A(\lambda)$ are the Hankel transformations of $\Phi(r,z,t)$ and $A(r)$ respectively, given by

$$\Phi(r,z,t) = \int_0^\infty \Psi(\lambda,z,t)\, J_0(\lambda r)\, \lambda\, d\lambda \;, \tag{11}$$

$$A(\lambda) = \frac{\alpha P \eta}{4 k_s \pi} \exp(-a^2 \lambda^2 / 8) \;. \tag{12}$$

Here $J_0(\lambda r)$ is the zero-order Bessel function of the first kind , and $\lambda$ is integration variable. Then, the Green's function method is used to solve Eqs.(8)-(10). In this case, the solution of Green's function corresponds to a solution of $\psi$ at a specific moment $(t = \tau)$ and the solution of the differential equations can be obtained conveniently. We define the Green's function by

$$\Psi(r,z,t) = \int_{-\infty}^{+\infty} Q(\tau)\, G(\lambda,z,t,\tau)\, d\tau \;, \tag{13}$$

where

$$Q(\tau) = \begin{cases} \exp(i\omega\tau) & \tau \geq 0 \\ 0 & \tau < 0 \end{cases} \tag{14}$$

Therefore, Eqs.(8)-(10) become , respectively

$$\left[-\lambda^2 + \frac{\partial^2}{\partial z^2} - \frac{1}{D_f}\frac{\partial}{\partial t}\right] G_f = 0 \qquad\qquad 0 \leq z \leq l_f \tag{15}$$

$$\left[-\lambda^2 + \frac{\partial^2}{\partial z^2} - \frac{1}{D_s}\frac{\partial}{\partial t}\right] G_s = -A(\lambda)\, e^{\alpha z}\, \delta(t-\tau) \qquad -l \leq z \leq 0 \tag{16}$$

$$\left[-\lambda^2 + \frac{\partial^2}{\partial z^2} - \frac{1}{D_b}\frac{\partial}{\partial t}\right] G_b = 0 \qquad\qquad -l-l_b \leq z \leq -l \tag{17}$$

Next, we use Laplace transformation to reduce the partial differential equations (15)-

(17) to ordinary ones. Defining the Laplace transform of the Green's function by
$$G_L(\lambda, z, p, \tau) = \frac{1}{2\pi} \int_0^\infty G(\lambda, z, t, \tau) e^{pt} dt, \tag{18}$$
we have
$$\left[-\lambda^2 + \frac{d^2}{dz^2} - \frac{p}{D_f}\right] G_{fL} = 0 \qquad\qquad 0 \leq z \leq l_f \tag{19}$$

$$\left[-\lambda^2 + \frac{d^2}{dz^2} - \frac{p}{D_s}\right] G_{sL} = -A(\lambda) e^{\alpha z} e^{-p\tau} \qquad -l \leq z \leq 0 \tag{20}$$

$$\left[-\lambda^2 + \frac{d^2}{dz^2} - \frac{p}{D_b}\right] G_{bL} = 0 \qquad\qquad -l - l_b \leq z \leq -l \tag{21}$$

The general solutions of these equations can be written as
$$G_{fL} = C(\lambda, p) e^{\beta_f z} + R(\lambda, p) e^{-\beta_f z}, \tag{22}$$
$$G_{sL} = U(\lambda, p) e^{\beta_s z} + V(\lambda, p) e^{-\beta_s z} - E(\lambda, p) e^{\alpha z}, \tag{23}$$
$$G_{bL} = W(\lambda, p) e^{\beta_b(z+l)} + D(\lambda, p) e^{-\beta_b(z+l)}, \tag{24}$$

where
$$E(\lambda, p) = \frac{A(\lambda)}{\alpha^2 - (\lambda^2 + p/D_s)} \exp(-p\tau), \tag{25}$$
and
$$\beta_j = \sqrt{(\lambda^2 + p/D_j)} \quad ; \quad j = f \text{ (fluid)}, \ s = \text{(sample)}, \ b = \text{(backing)}. \tag{26}$$

The coefficients U, V, W, R, C, D are determined by using the boundary conditions (5)-(7). We get
$$U(\lambda, p) = \frac{(1+b)(s+g)e^{\beta_s l} - (s-b)(1-g)e^{-\alpha l}}{H(\lambda, p)} E(\lambda, p), \tag{27a}$$

$$V(\lambda, p) = \frac{(1+g)(b-s)e^{-\alpha l} + (g+s)(1-b)e^{-\beta_s l}}{H(\lambda, p)} E(\lambda, p), \tag{27b}$$

$$W(\lambda, p) = U(\lambda, p) e^{-\beta_s l} + V(\lambda, p) e^{\beta_s l} - E(\lambda, p) e^{-\alpha l}, \tag{27c}$$
$$R(\lambda, p) = U(\lambda, p) + V(\lambda, p) - E(\lambda, p), \tag{27d}$$
$$H(\lambda, p) = (1+g)(1+b)e^{\beta_s l} - (1-g)(1-b)e^{-\beta_s l}, \tag{27e}$$
$$C(\lambda, p) = D(\lambda, p) = 0, \tag{27f}$$

Where
$$g = \frac{k_f \beta_f}{k_s \beta_s}, \quad b = \frac{k_b \beta_b}{k_s \beta_s}, \quad s = \frac{\alpha}{\beta_s}.$$

Finally, the temperature distributions for different regions are obtained by taking the inverse Laplace transform and then the inverse Hankel transform of Eqs.(22-24). We find
$$\Phi_f(r, z, t) = \int_0^\infty R(\lambda, \omega) e^{-\beta_f z} J_0(\lambda r) e^{i\omega t} \lambda d\lambda, \tag{28}$$

$$\Phi_s(r, z, t) = \int_0^\infty \left[U(\lambda, \omega) e^{\beta_s z} + V(\lambda, \omega) e^{-\beta_s z} - E(\lambda, \omega) e^{-\alpha z}\right] J_0(\lambda r) \lambda e^{i\omega t} d\lambda, \tag{29}$$

$$\Phi_b(r,z,t) = \int_0^\infty W(\lambda,\omega)\, e^{\beta_b(z+l)} J_0(\lambda r)\, e^{i\omega t} \lambda d\lambda. \tag{30}$$

In these equations the coefficients U, V, W, and R are given by Eqs.(27a-d) respectively, except p is replaced by $i\omega$. We also redefine $\beta_j$ as

$$\beta_j = \sqrt{(\lambda^2 + i\omega/D_j)} \quad ; \quad j = f\text{ (fluid)}, \text{ s (sample)}, \text{ b = (backing)} \tag{31}$$

For determination of the photothermal signal, it is $\Phi_f(r,z,t)$ that is important. For z=0, as we would expect

$$\Phi_f(r,0,t) = \Phi_s(r,0,t) \equiv \Psi_s(\lambda,0,t). \tag{32}$$

Therefore, Eq.(28) can be written as

$$\Phi_f(r,z,t) = \int_0^\infty \Psi_s(\lambda,0,t)\, e^{-\beta_f z} J_0(\lambda r)\, e^{i\omega t} \lambda d\lambda. \tag{33}$$

The observable temperatures are just the real parts of $\Phi(r,z,t)$. Let us denote the real part of $\Phi_f$ by $T_f$, the real and imaginary parts of $\Psi_s(\lambda,0,t)$ [which is the same as $R(\lambda,\omega)$] by $R_1$ and $R_2$ and the real and imaginary parts of $\beta_f$ by $\beta_{f_1}$ and $\beta_{f_2}$, respectively. Then $T_f(r,z,t)$ may be written as

$$T_f(r,z,t) = \int_0^\infty |\Psi_s(\lambda,0,t)|\, e^{-\beta_{f_1} z} J_0(\lambda r)\, \lambda \sin(\beta_{f_2} z - \omega t + \delta) d\lambda, \tag{34}$$

where $\delta = \tan^{-1}(R_2/R_1)$ and $|\Psi_s(\lambda,0,t)| = \sqrt{R_1^2 + R_2^2}$. The expression (34) has a simple interpretation. At z =0, $T_f$ is equal to the surface temperature of the sample $T_s$. With increasing z, $T_f$ behaves like a thermal wave with exponentially decaying amplitude and period $T = 2\pi/\omega$. The thermal length $\sigma_f$ and the wavelength $\lambda_f$ are given by

$$\sigma_f = \frac{1}{\beta_{f_1}} = \frac{1}{\text{Re}\sqrt{\lambda^2 + i\omega/D_f}}, \tag{35}$$

$$. \tag{36}$$

By using Eqs.(27a-e) and (12) one may rewrite Eq.(34) as

$$T_f(r,z,t) = \frac{\alpha P \eta}{4\pi k_s} \int_0^\infty e^{-\beta_{f_1} z} J_0(\lambda r)\, \lambda \sin(\beta_{f_2} z - \omega t + \delta) \times$$

$$\left| \frac{-(1+b)(1-s)e^{\beta_s l} + (1-b)(1+s)e^{-\beta_s l} - 2(s-b)e^{-\alpha l}}{(1+b)(1+g)e^{\beta_s l} - (1-b)(1-g)e^{-\beta_s l}} \cdot \frac{e^{-a^2\lambda^2/8}}{\alpha^2 - \beta_s^2} \right| d\lambda. \tag{37}$$

Since Eq.(37) is not in a closed form, it should be evaluated numerically. A useful special case occurs for laser beams of very large diameters. In this case $\lambda$ approaches zero, so $A(\lambda)$ and as a result $E(\lambda)$ behave as delta function [cf. Eqs.(12),(25)]. In this limit, Eqs.(28)-(30) can be rewritten as following

$$\Phi_f(z,t) = R(\omega)\exp(-\beta_f z)\exp(i\omega t), \tag{38}$$

$$\Phi_s(z,t) = [U(\omega)\exp(\beta_s z) + V(\omega)\exp(-\beta_s z) - E(\omega)\exp(-\alpha z)]\exp(i\omega t), \tag{39}$$

$$\Phi_b(z,t) = W(\omega)\exp[\beta_b(z+l)]\exp(i\omega t). \tag{40}$$

Now, these relations are in closed form. This model is referred to as the 1-D model, because the diffusion of the heat occurs only in one dimension, that is, in the z-direction. The theory for the 1-D model was first developed by Rosencwaig and Gersho[17]. In 1-D model, thermal length and thermal wavelength [Eqs.(35) and (36)] reduces respectively to the forms,

$$\bar{\sigma}_f = \frac{1}{\mathrm{Re}\beta_f} = \sqrt{\frac{2D_f}{\omega}} \quad , \tag{41}$$

$$\bar{\lambda}_f = \frac{2\pi}{\mathrm{Im}\beta_f} = 2\pi\sqrt{\frac{2D_f}{\omega}} . \tag{42}$$

Similar simple interpretation can also be given to $\Phi_b$ and $\Phi_s$. The temperature of the backing $\Phi_b$ is represented by backward traveling thermal waves. The temperature of the sample itself is represented by a forward traveling and a backward traveling thermal wave, and a term representing the absorption of the laser energy. It is instructive to consider the typical values of $\bar{\sigma}_f$ and $\bar{\sigma}_s$ as they set the scale over which various physical effects are observable. For $N_2$ at atmospheric pressure, $\bar{\sigma}_f$=0.85mm for f=10Hz and it is 0.27mm for f=100Hz. For a sample of α-Si:H, $\bar{\sigma}_s$=0.18mm for f=10Hz and it is 0.06mm for f=100Hz.

It is also useful to determine the peak value of $T_f$. For this purpose we rewrite Eq.(34) as

$$T_f(r,z,t) = \left\{\int_0^\infty \left[R_2 \cos\beta_{f_2} z + R_1 \sin\beta_{f_2} z\right]\exp(-\beta_{f_1} z) J_0(\lambda r)\lambda d\lambda\right\}\cos\omega t$$

$$- \left\{\int_0^\infty \left[R_1 \cos\beta_{f_2} z - R_2 \sin\beta_{f_2} z\right]\exp(-\beta_{f_1} z) J_0(\lambda r)\lambda d\lambda\right\}\sin\omega t . \tag{43}$$

In this expression, the two terms represent, respectively, the in-phase and quadrature components of the temperature. They can be measured individually by phase sensitive detection techniques. The peak value, then is simply

$$T_{fo}(r,z,t) = \left\{\left[\int_0^\infty \left[R_2 \cos\beta_{f_2} z + R_1 \sin\beta_{f_2} z\right]\exp(-\beta_{f_1} z) J_0(\lambda r)\lambda d\lambda\right]^2 \right.$$

$$\left. + \left[\int_0^\infty \left[R_1 \cos\beta_{f_2} z - R_2 \sin\beta_{f_2} z\right]\exp(-\beta_{f_1} z) J_0(\lambda r)\lambda d\lambda\right]^2\right\}^{1/2} . \tag{44}$$

**2nd. Photothermal Deflection**

In this section the temperature distribution $T_f$, Eq.(37), will be used to calculate the photothermal deflection of a probe beam propagating through the fluid in thermal contact with the sample. Figure2 shows an illustration of a deflection experiment. The pump beam is incident on the sample in the z-direction. The sample itself is in the x-y plane and the probe beam propagates in the x-direction.

The temperature distribution gives rise to a spatially varying index of refraction given by

$$n(\vec{r},t) = n_0 + \frac{\partial n}{\partial T}T(\vec{r},t) , \tag{45}$$

where $n_0$ is the index of refraction of the medium at ambient temperature. The deflection of a beam propagating through such a spatially varying index of refraction can be found from Fermat's principle, which states that the optical path length is a minimum of the system Hamiltonian. The result is [17]

$$\frac{d}{ds}\left(n_0 \frac{d\vec{r}_0}{ds}\right) = \vec{\nabla}_\perp n(\vec{r},t), \qquad (46)$$

where $ds$ and $d\vec{r}_0$ represent the beam path and beam deflection, respectively, as shown in Fig. 3. Furthermore, $\vec{\nabla}_\perp$ denotes the gradient normal to the beam path $ds$. Combining the relations (45) and (46) results in

$$\frac{d\vec{r}_0}{ds} = \frac{1}{n_0}\frac{\partial n}{\partial T}\int_{path} \vec{\nabla}_\perp T(\vec{r},t)\,ds. \qquad (47)$$

Figure 3 shows that the deflection angle $\frac{d\vec{r}_0}{ds}$ has two components, a tangential deflection $\theta_t$, across the sample surface, and a normal deflection $\theta_n$, perpendicular to the sample surface, given respectively by

$$dr_t/ds = \sin\theta_t \approx \theta_t \quad , \quad dr_n/ds = \sin\theta_n \approx \theta_n.$$

The deflection components are found from

$$\theta_t = \frac{1}{n_0}\frac{\partial n}{\partial T}\int_{-\infty}^{\infty}\frac{\partial T_f}{\partial y}dx, \qquad (48)$$

$$\theta_n = \frac{1}{n_0}\frac{\partial n}{\partial T}\int_{-\infty}^{\infty}\frac{\partial T_f}{\partial z}dx. \qquad (49)$$

The spatial dependence of $T_f$ is simple enough that the integration over $dx$ can be carried out in closed form. Substituting Eq.(34) into the tangential deflection expreesion (48) gives

$$\theta_t = \frac{1}{n_0}\frac{\partial n}{\partial T}\int_0^\infty |\psi_s(\lambda,0,t)|\exp(-\beta_{f_1}z)\lambda\sin(\beta_{f_2}z-\omega t+\delta)d\lambda \int_{-\infty}^{\infty}\frac{y}{r}\frac{\partial}{\partial r}J_0(\lambda r)dx$$

(50)

The integration over $dx$ can be done by making the substitutions $dx = (r/x)dr$ and then $\upsilon = r/y$. Therefore the integral is transformed into

$$I_t = \int_{-\infty}^{\infty}\frac{y}{r}\frac{\partial}{\partial r}J_0(\lambda r)dx = -2\lambda\int_1^\infty\frac{J_1(\lambda y\upsilon)}{\sqrt{\upsilon^2-1}}d\upsilon,$$

which evaluates to [19]

$$I_t = -2\lambda y(-\pi/2)J_{1/2}(\lambda y/2)N_{1/2}(\lambda y/2) = -2\sin(\lambda y).$$

Thus the tangential deflection (50) can be written as

$$\theta_t = -\frac{2}{n_0}\frac{\partial n}{\partial T}\int_0^\infty |\psi_s(\lambda,0,t)|\exp(-\beta_{f_1}z)\lambda\sin(\beta_{f_2}z-\omega t+\delta)\sin(\lambda y)d\lambda. \qquad (51)$$

The normal deflection is calculated in a similar manner. The final result is given by

$$\theta_n = \frac{2}{n_0}\frac{\partial n}{\partial T}\int_0^\infty |\psi_s(\lambda,0,t)|\exp(-\beta_{f_1}z)\{\beta_{f_1}\cos(\beta_{f_2}z-\omega t+\delta)+\beta_{f_2}\sin(\beta_{f_2}z-\omega t+\delta)\}\cos(\lambda y)d\lambda$$

(52)

Generally the experiments are performed using the lock-in techniques. In this case the peak value of $\theta_t$ and $\theta_n$ are observed. These peak values can be found from Eqs.(51) and (52) to be

$$\theta_{t0} = \frac{2}{n_0}\frac{\partial n}{\partial T}\frac{1}{\sqrt{2}}\left\{\left(\int_0^\infty \exp(-\beta_{f_1} z)\lambda \sin(\lambda y)D_0 d\lambda\right)^2 + \left(\int_0^\infty \exp(-\beta_{f_1} z)\lambda \sin(\lambda y)E_0 d\lambda\right)^2\right\}^{1/2},$$
(53)

$$\theta_{n0} = \frac{2}{n_0}\frac{\partial n}{\partial T}\frac{1}{\sqrt{2}}\left\{\left(\int_0^\infty \exp(-\beta_{f_1} z)\cos(\lambda y)A_0 d\lambda\right)^2 + \left(\int_0^\infty \exp(-\beta_{f_1} z)\cos(\lambda y)B_0 d\lambda\right)^2\right\}^{1/2},$$
(54)

where

$$A_0 = (R_1\beta_{f_1} - R_2\beta_{f_2})\cos(\beta_{f_2} z) + (R_2\beta_{f_1} + R_1\beta_{f_2})\sin(\beta_{f_2} z),$$
$$B_0 = (R_1\beta_{f_1} - R_2\beta_{f_2})\sin(\beta_{f_2} z) - (R_2\beta_{f_1} + R_1\beta_{f_2})\cos(\beta_{f_2} z),$$
$$D_0 = R_2\cos(\beta_{f_2} z) - R_1\sin(\beta_{f_2} z),$$
$$E_0 = R_1\cos(\beta_{f_2} z) + R_2\sin(\beta_{f_2} z).$$

### III. Numerical Results and Discussions

In this section we will describe the results obtained for the temperature distribution within the three layer system and photothermal deflection of the probe beam. The temperature profile $T_f$ [Eq.(37)] and the thermal deflection signals $\theta_t$ [Eq.(51)] and $\theta_n$ [Eq.(52)] can not be evaluated in closed form, so numerical methods must be used.

We first consider the temperature distribution $T_f$. It is useful to look at the integrand in Eq.(37). The Bessel function contributes to oscillation in the integrand, while the exponential term $\exp(-a^2\lambda^2/8)$ makes the function damp out. This is rather convenient, because we can replace the upper limit on the integration by a certain value $\lambda_m$, such that the integrand reduces to negligible values for $\lambda > \lambda_m$. We choose $\lambda_m$ to be such that to make $\exp(-a^2\lambda^2/8) < \exp(-2)$. Note that $\lambda_m$ is inversely proportional to $a$. Depending on the values of $\omega$ and $a$, the range of integration 0 to $\lambda_m$ is divided into 3 to 10 regions. Each region has been calculated by Gaussian quadrature of 64-points. In the following, the backing and fluid are assumed to be Corning glass ($k_b$=1W/m.K, $D_b$=6×10$^{-7}$m$^2$/s) and $N_2$ gas ($k_f$=2.6×10$^{-2}$W/m.K, $D_f$ = 23 ×10$^{-6}$ m$^2$/s ) [20], respectively. It is also assumed that the pump beam power P=1W and light-heat conversion coefficient $\eta$=1.

In figure 4a, assuming that the solid sample to be Ge, we have plotted the temperature ( temperature deviation from the ambient value) distribution at the surface of the sample $T_s(r,0,t)$ [= $T_f(r,0,t)$] as a function of r for several values of time in one modulation cycle. Equation (37) has been evaluated by using the parameters given on the figure. It is found that at $\omega t = 2\pi\times 10t \approx 9.5\times 10^{-3}\pi$, $T_f(0,0,t)=0$. This value of $\omega t$, which is denoted by $\theta$, is used as a reference in plotting other curves in figure 4. Curves at $\omega t = \theta + \pi/4$, $\omega t = \theta + \pi/2$, $\omega t = \theta + 3\pi/4$ and $\omega t = \theta + 3\pi/2$ are shown, and they show expected behavior with respect to time. The foot print of the heat essentially follows the spatial profile of the pump beam because the diffusion length $\sigma_s \approx 2.5\times 10^{-2}$ mm is much smaller than the beam radius a . This means that the heat does not diffuse very much beyond the extent of the pump beam in the x-y plane. It is important to note that the temperature fluctuates between positive and negative values , because we have calculated only the oscillatory part of the temperature. The actual temperature consists of the oscillatory part superimposed on a time independent part ( contributed by the factor A(r) exp($\alpha$z) in the source term), which we have not calculated. In figure 4b , we have plotted the temperature distribution $T_s(r,0,t)$ as a function of r for f=100Hz . Other parameters are the same as those in Fig.4a . As it is seen the general behavior of $T_s$ is not so sensitive to the change of modulation frequency in such a manner that the difference between two figures is rather quantitative than qualitative.

Figures 5a and 5b show the temperature distribution $T_f(0,z,t)$ as a function of z for several values of time in one modulation cycle and for f = 10 Hz and f= 100 Hz , respectively . The surface temperature is sinousoidally modulated as found in Fig.4 , and a thermal wave propagate in the fluid . The thermal wave is strongly attenuated with the decay length of the order of $\sigma_f$ .

Figure 6 shows the peak values of the temperature in the fluid $T_{f0}$ as a function of the distance z for three different values of the modulation frequency f . Two effects should be noted . First , the temperature of the sample surface decreases with increasing modulation frequency because of the thermal inertia of the sample.In other words , the sample is unable to respond to the intensity changes to a lesser and lesser degree as the modulation frequency of the pump laser increases. Second , the effective thermal length $\sigma_f$ decreases with increasing frequency , thereby making the decay of photothermal signal with z faster .

Figure 7a shows the dependence of the peak values of the temperature at sample surface on r for three different types of solid samples . As the thermal diffusivity $D_s$ increases the temperature decreases because the heat is able to diffuse further . Moreover , the profile of the temperature distribution gets broader with increasing $D_s$ . Here the optical absorption coefficient $\alpha$ for each of the three samples is much larger than the corresponding values of $\bar{\sigma}_s^{-1}$ . This case corresponds to the situation when most of the laser energy is absorbed near the surface of the sample. It appears that the thermal diffusion in the negative z-direction dominates over thermal diffusion in the r-direction in this case. We also find no significant effect of the thermal diffusivity of the fluid on the temperature profiles at the surface. In Fig. 7b we have plotted $T_{f0}$ at z=0 as a function of r . Here the modulation frequency is assumed to be f = 100Hz. Other parameters are the same as those in Fig.7a . Comparison with Fig .7a reveals that $T_{f0}$ decreases with increasing f . In fact as the modulation frequency increases the transmission coefficient of thermal wave at the boundary of sample and fluid increases and in consequence $T_{f0}$ decreases . In addition with increasing f , the profile temperature distribution gets narrower , as expected .

Figure 8 shows the temperature profile as a function of r in the fluid for the parameters shown on the figure . The temperature profile is seen to broaden with increasing z , as expected .

We now proceed to evaluate numerically the deflection signal [Eqs.(51),(52),(53) and (54)] . For this purpose , as before , we have used Gaussian quadrature of 64-points. The deflection takes place in three dimensions ; the probe beam propagates in the x-direction and is deflected normally away from the sample surface into the z-direction , $\theta_n$ in Eq.(52), and tangential to the sample surface into the y-direction, $\theta_t$ in Eq.(51), as was shown in Figs.1 and 3 . As before , we assume that the fluid and backing are nitrogen gas and Corning glass , respectively . It is also assumed that pump power P=1W , $\partial n / \partial T = 9.4 \times 10^{-7} K^{-1}$ (at room temperature) and $\eta$=1 .

For the glass-Ge-nitrogen system , Figures 9a and 9b give $\theta_t$ and $\theta_n$ , respectively, as functions of y at z=0 for different values of time in one modulation cycle and for f=10 Hz. As before , the initial value of the time , $\theta = \omega t$ is chosen such that $T_f$=0 at y=0 at this time . The normal deflection is maximum at y = 0 but in the tangential deflection there is no signal when the pump is centered on the probe at y= 0 , the probe beam is pulled equally in each direction . To either side of this point , the probe beam is deflected in opposite directions , up or down , thus the change in sign on either side. In both figures the distribution is reflective of the Gaussian pump profile.

In figures 10a and 10b the peak values of the tangential and normal photothermal deflection are plotted against the y coordinate , for several different values of z and for f = 10Hz. These are the signals that are generally measured using the lock-in techniques. As the distance from the surface increases the deflection signal intensity decreases and the width increases since the heat is dispersed throughout more of the fluid. It is interesting to note that the gradient of $\theta_{n0}$, for the values of the parameters chosen here, is ~2 orders of magnitudes smaller than that that of $\theta_{t0}$ . In fact the gradient of photothermal deflection ( curvature of refractive index ) characterizes the inverse of the focal length of the thermal lens [21] that is produced by the heating action of the Gaussian laser beam. Therefore we find that the peak value of the inverse focal length of the photothermal lens in the z direction is ~2 orders of magnitude smaller than that of in y direction .

The effect of changing the sample diffusivity/conductivity on the tangential deflection signal is shown in Fig.11. The peak value decreases as the sample diffusivity is increased. This is the behavior that was seen in the fluid temperature of figures 7a and 7b. As modeled the absorption coefficient of the sample is very large and the radiation absorption is taking place at the surface. The heat then diffuses preferentially into and throughout the sample due to the relatively low thermal conductivity of the fluid. For a small sample diffusivity the signal is larger due to the heat lingering at the surface for a longer time , allowing more heat to conduct into and through the fluid. The increased heat results in a larger index gradient and larger deflection. With a larger sample diffusivity/conductivity the heat quickly disperses throughout the sample, leaving only the initial surface heat to diffuse into the fluid . This results in a decrease in the signal intensity and slight increase in the signal width. The plot of peak value of normal deflection signal as a function of y for different values of diffusivity/ conductivity (not shown) also reveals similar dependence on $D_s$/ $k_s$ as the peak value of tangential deflection signal. The peak value of normal deflection is greater than that of the tangential deflection due to the increased distance from heating epicenter. Furthermore, its gradient is much smaller than that of tangential deflection. This shows that irrespective of the sample diffusivity/conductivity the peak value of the inverse focal length of the photothermal lens in the z direction is much smaller than that of in y direction .

Figure 12 shows the effect of the modulation frequency on the tangential deflection signal. The signal and its width decrease with increasing frequency , as expected. Decreasing the signal width with increasing modulation frequency shows that for larger frequency the focal length of photothermal lens in y direction decreases . The plot of peak value of normal deflection signal as a function of y for different values of modulation frequency (not shown) also reveals similar dependence on frequency as the peak value of tangential deflection signal.

**IV. Conclusions**

We have presented a detailed theoretical description of the three dimensional photothermal deflection, induced by modulated cw laser excitation, for a three layer system consisting of a transparent fluid , an optically absorbing solid sample and a backing material. Some of the important results are the following : (i) the laser induced temperature of the sample surface decreases with increasing modulation frequency of the pump laser. (ii) The effective thermal length decreases with increasing modulation frequency , thereby making the decay of photothermal signal with z faster. (iii) As the modulation frequency increases the temperature distribution $T_{f0}$ decreases and gets narrower. (iv) As the distance from the surface of solid sample increases the deflection signal intensity decreases and its width increases. (v) The focal length of the photothermal lens , produced by the heating action of the pump laser , in the z direction is much greater than that of in y direction. (vi) The increasing of diffusivity/ conductivity of solid sample results in a decrease in the deflection signal intensity and slight increase in the signal width. (vii) The normal deflection is greater than the tangential deflection , while its gradient is much smaller than that of tangential deflection. (viii) The deflection signal and its width decrease with increasing modulation frequency.

**References**

[1] A. Rosencwaig, *Photoacoustics and photoacoustic spectroscopy*, (John Wiley, NY, 1980).

[2] D.P. Almond and P. M. Patel, *Photothermal Science and Techniques*, Chapman&Hall, (1996); A. Mandelis, Physics Today, Vol.**53**, No.8, 29 (2000).

[3] M. Luukkala, in Scanned Image Microscopy, E. A. Ash. Ed.(Academic, London, 1980).

[4] W. B. Jackson, N. M. Amer, A. C. Boccara, and D. Fournier, Appl. Opt.**20**, 1333 (1981).

[5] J. C. Murphy and L. C. Aamodt, Appl. Phys. Lett. **38**, 196 (1981).

[6] S. Ameri, E. Ash, V. Neuman, and C. R. Petts, Electron Lett. **17**, 337 (1981).

[7] M. A. Olmstead, S. E. Kohn, and N. M. Amer, Bull. Am. Phys. Soc **27**, 227 (1982).

[8] J. Opsal, A. Rosencwaig, and D. L. Willenburg, Appl. Opt.**22**, 3169 (1983).

[9] L. C. M. Miranda, Appl. Opt.**22**, 2882 (1983).

[10] M. A. Olmstead, N. M. Amer, S. Kohn, D. Fournier, and A. C. Boccara, Appl. Phys. A**32**, 141 (1983).

[11] M. A. Olmstead and N. M. Amer, J. Vac. Sci&Technol. B1, 751 (1983).

[12] N. Y. Yacoubi, B. Girault, and J. Fesquet, Appl. Opt.**25**, 4622 (1986).

[13] M. Soltanolkotabi, G. L. Bennis, and R. Gupta, J. Appl. Phys.**85**(2), 794, .

[14] B. C. Li, J. Appl. Phys.**68**, 482 (1990).

[15] J.-C. Cheng, L. Wu, and S.-Y. Zhang, J. Appl. Phys.**76**, 716 (1994).

[16] J.-C. Cheng and S.-Y. Zhang, J. Appl. Phys. **74**, 5718 (1993).

[17] A. Rosencwaig and A. Gersho, J. Appl. Phys.**47**, 64 (1976).

[18] M. Born and E. Wolf, *Principles of Optics*, (Pergamon Press, Oxford, 1970).

[19] I. M. Ryzhik, Alan Jeffery, and I. S. Gradshteyn, *Table of Integrals, Series, and Products*,(Academic Press, San Diego, 1994).

[20] CRC Handbook of Chemistry and Physics, 78[th] ed., CRC Press (1997).

[21] H. L. Fang and R. L. Swofford ," *The Thermal Lens in Absorption Spectroscopy",* in Ultrasensitive Laser Spectroscopy, Ed. By D. S. Kliger 1983, Academic press Inc (London) LTD, pp175-232.

**Figure Captions**

**Fig.1** Geometry of the three layer system of photothermal deflection effect. Each region is taken to be of infinite extent in the x-y plane.

**Fig.2** An illustration of the photothermal deflection spectroscopy.

**Fig.3** Probe beam deflection normal and tangential to the sample surface. The box is within the fluid region, with the sample surface parallel to the nearest box face.

**Fig.4a** Surface temperature $T_s(r,0,t)$ as a function of r for five different times in one modulation cycle and f=10Hz.

**Fig.4b** Surface temperature $T_s(r,0,t)$ as a function of r for five different times in one modulation cycle and f=100Hz.

**Fig.5a** Temperature distribution $T_f$ as a function of z at r=0 for different times and f= 10Hz.

**Fig.5b** Temperature distribution $T_f$ as a function of z at r=0 for different times and f= 100Hz.

**Fig.6** Temperature distribution $T_{f0}$ (peak value) as a function of z at r=0 for different values of modulation frequency.

**Fig.7a** Temperature distribution $T_{f0}$ (peak value) as a function of r at z=0 for three sample diffusivities and f=10Hz.

**Fig.7b** Temperature distribution $T_{f0}$ (peak value) as a function of r at z=0 for three sample diffusivities and f=100Hz.

**Fig.8** Temperature distribution $T_{f0}$ (peak value) as a function of r for different values of z and f=10Hz.

**Fig.9a** Transverse deflection $\theta_t$ as a function of y for three different times in one modulation cycle and f= 10Hz.

**Fig.9b** Normal deflection $\theta_n$ as a function of y for three different times in one modulation cycle and f= 10Hz.

**Fig. 10a** Transverse deflection $\theta_{t0}$ (peak value) as a function of y for three different values of z.

**Fig. 10b** Normal deflection $\theta_{n0}$ (peak value) as a function of y for three different values of z.

**Fig. 11** Transverse deflection $\theta_{t0}$ (peak value) as a function of y at z=0 and for three different values of sample diffusivity/conductivity.

**Fig. 12** Transverse deflection $\theta_{t0}$ (peak value) as a function of y at z=0 and for three different values of modulation frequency.

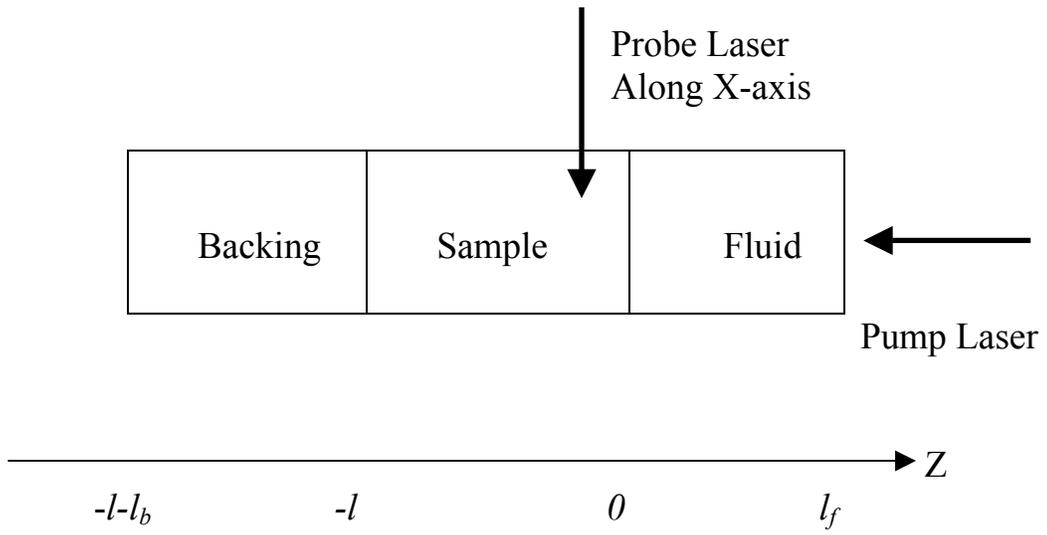

Fig.1. Geometry of the three layer system of photothermal deflection effect. Each region is taken to be of infinite extent in the xy-plane.

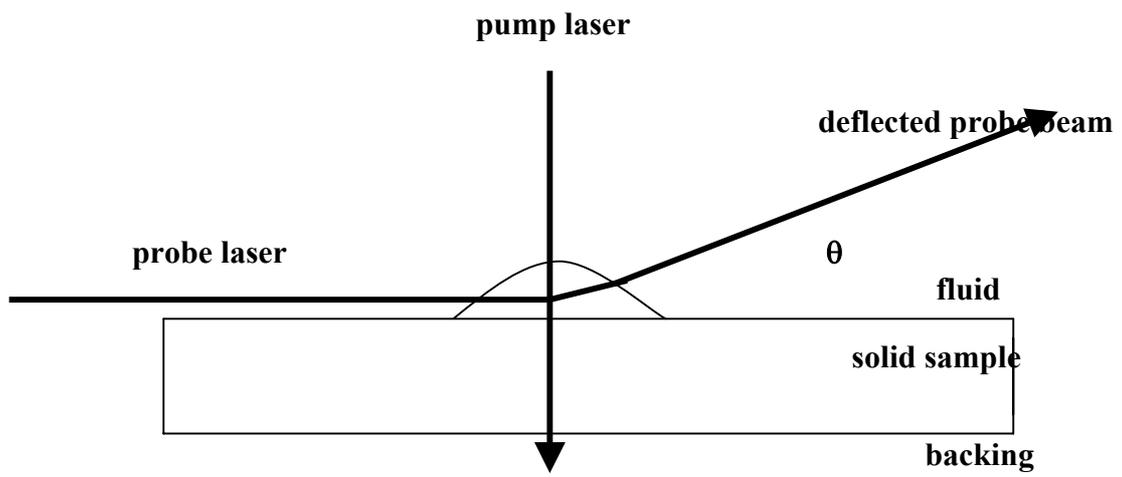

**Fig.2.** An illustration of the photothermal deflection spectroscopy

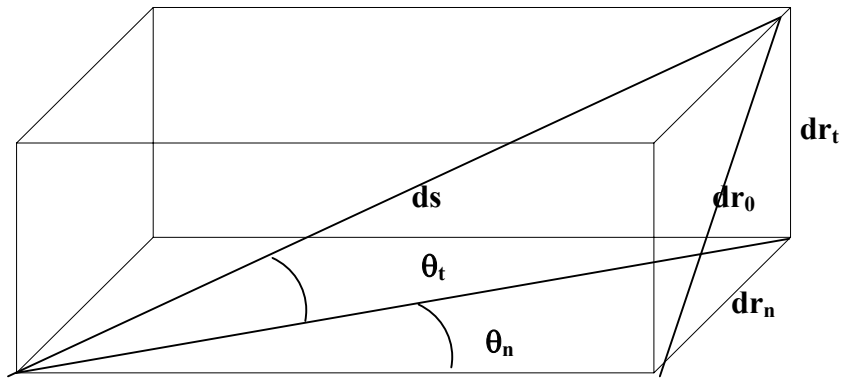

**Fig.3.** Probe beam deflection normal and tangential to the sample surface. The box is within the fluid region, with the sample surface parallel to the nearest box face.

Ge : $k_s = 64$ W/m.K , $\alpha = 5*10^7$ $m^{-1}$ , $d_s = 373*10^{-7}$ $m^2/s$

$z = 0$ , $f = 10$ Hz , $a = 1$ mm , $l = 1$ mm

a) $\omega t = \theta$ , b) $\omega t = \theta + \pi/4$ , c) $\omega t = \theta + \pi/2$
d) $\omega t = \theta + 3\pi/4$ , e) $\omega t = \theta + 3\pi/2$

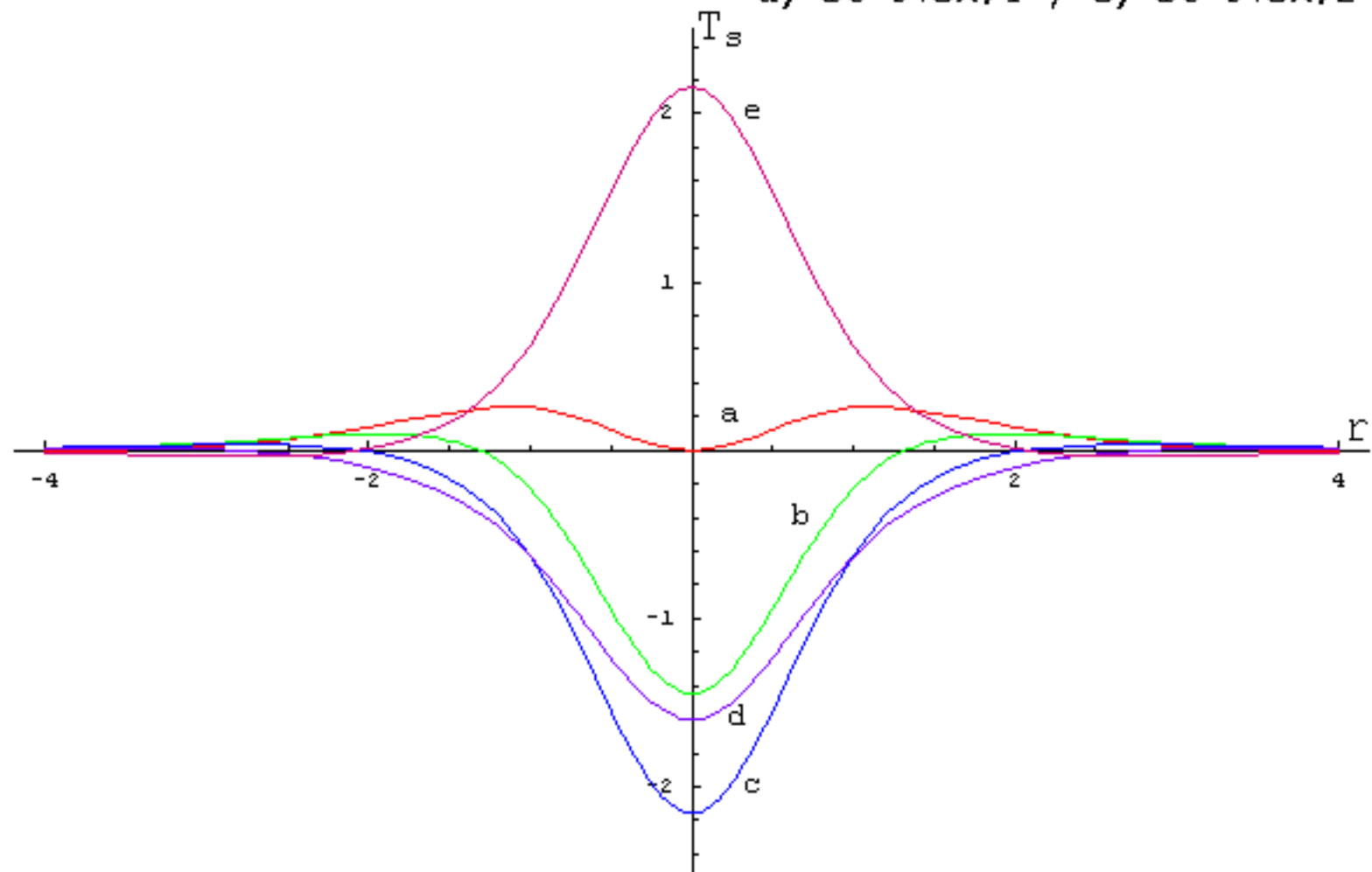

Ge : $k_s$=64 W/m.K , $\alpha=5*10^7$ $m^{-1}$ , $d_s=373*10^{-7}$ $m^2/s$

z=0 , f=100 Hz , a=1 mm , l= 1 mm

a) $\omega t=\theta$ , b) $\omega t=\theta+\pi/4$ , c) $\omega t=\theta+\pi/2$
d) $\omega t=\theta+3\pi/4$ , e) $\omega t=\theta+3\pi/2$

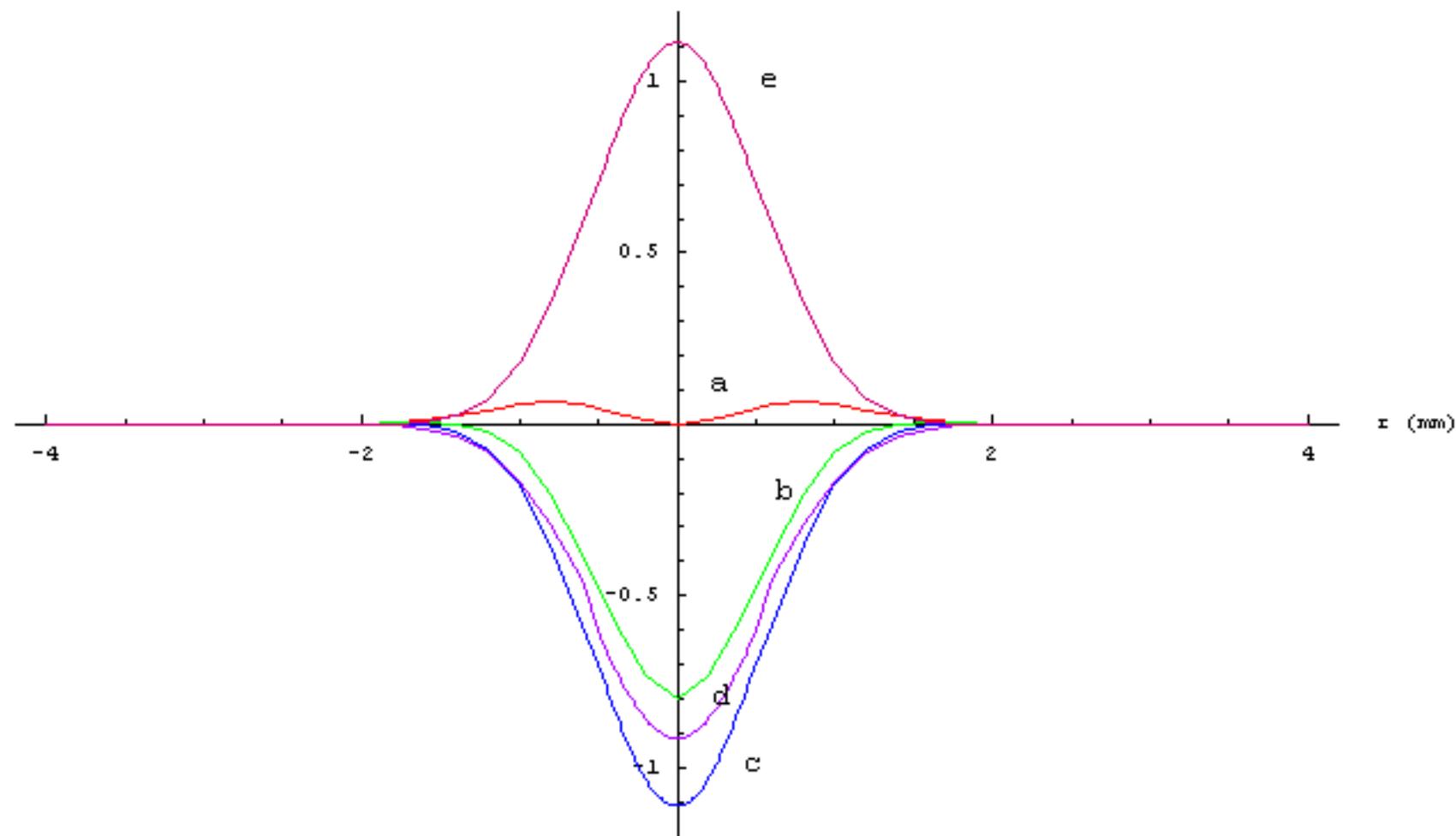

Ge : $k_s=64$ W/m.K , $\alpha=5*10^7$ $m^{-1}$ , $D_s=373*10^{-7}$ $m^2/s$

r=0 , f=10 Hz , a=1 mm , l= 1 mm

a) $\omega t=\theta$ , b) $\omega t=\theta+\pi/4$ , c) $\omega t=\theta+\pi/2$
d) $\omega t=\theta+3\pi/4$ , e) $\omega t=\theta+3\pi/2$

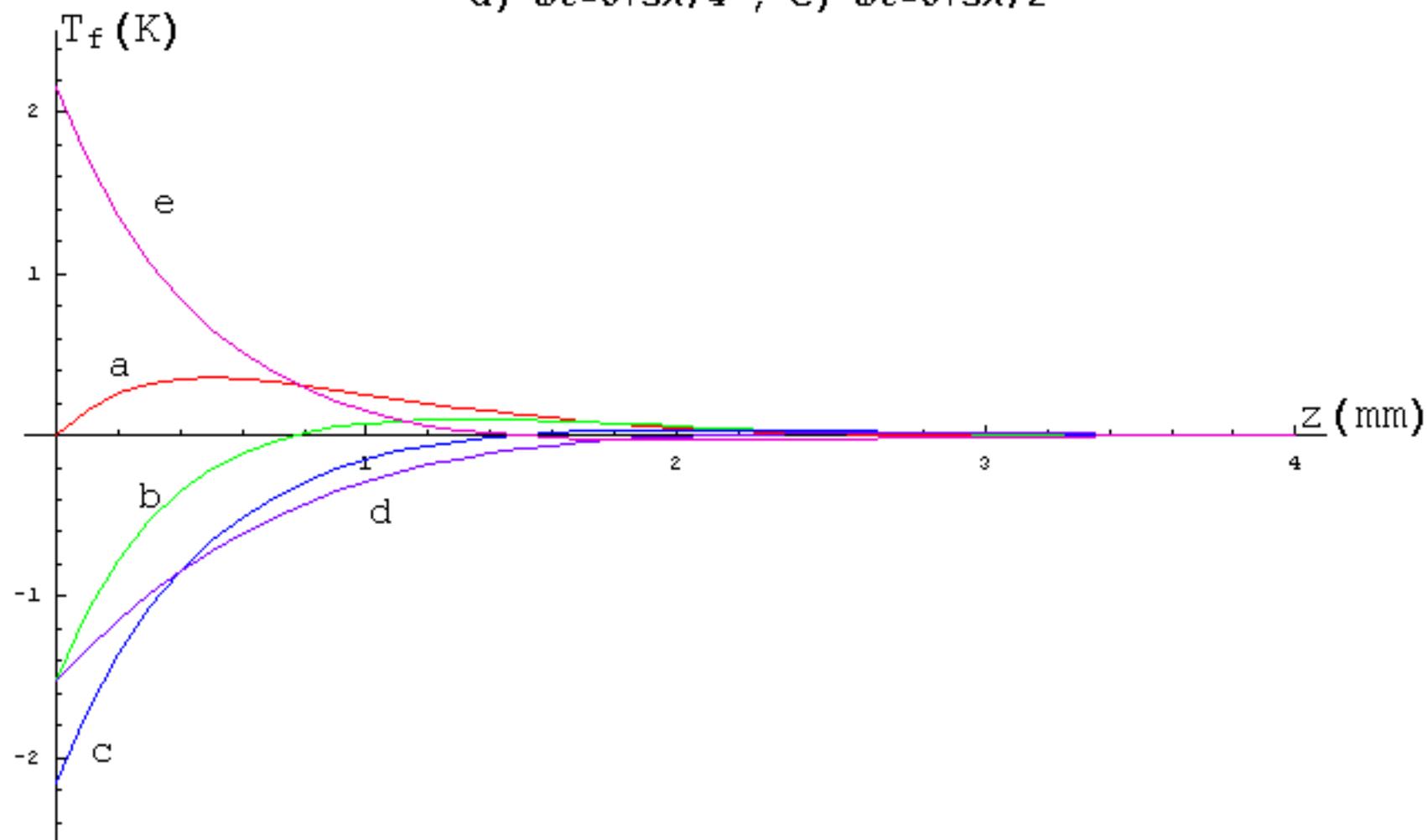

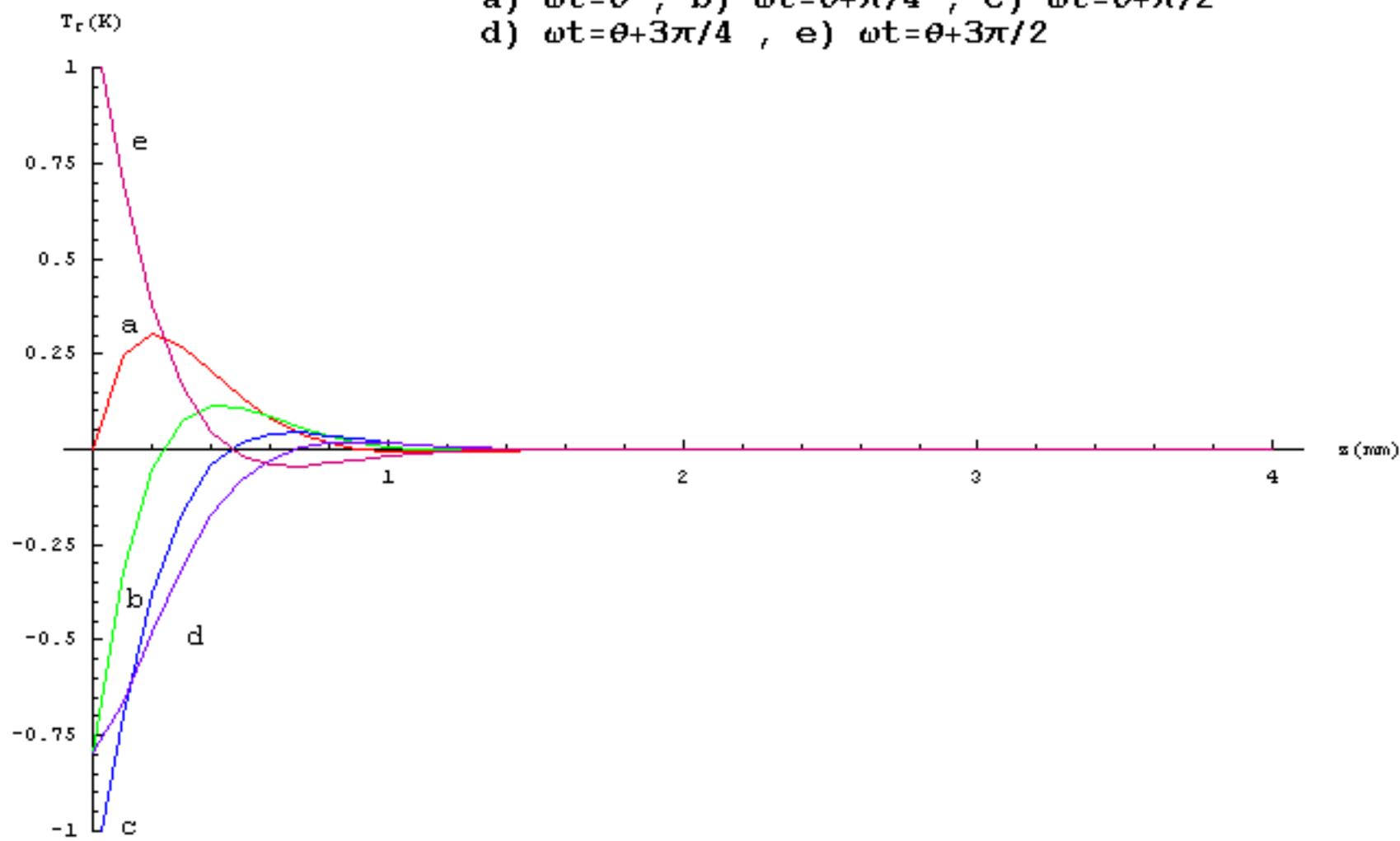

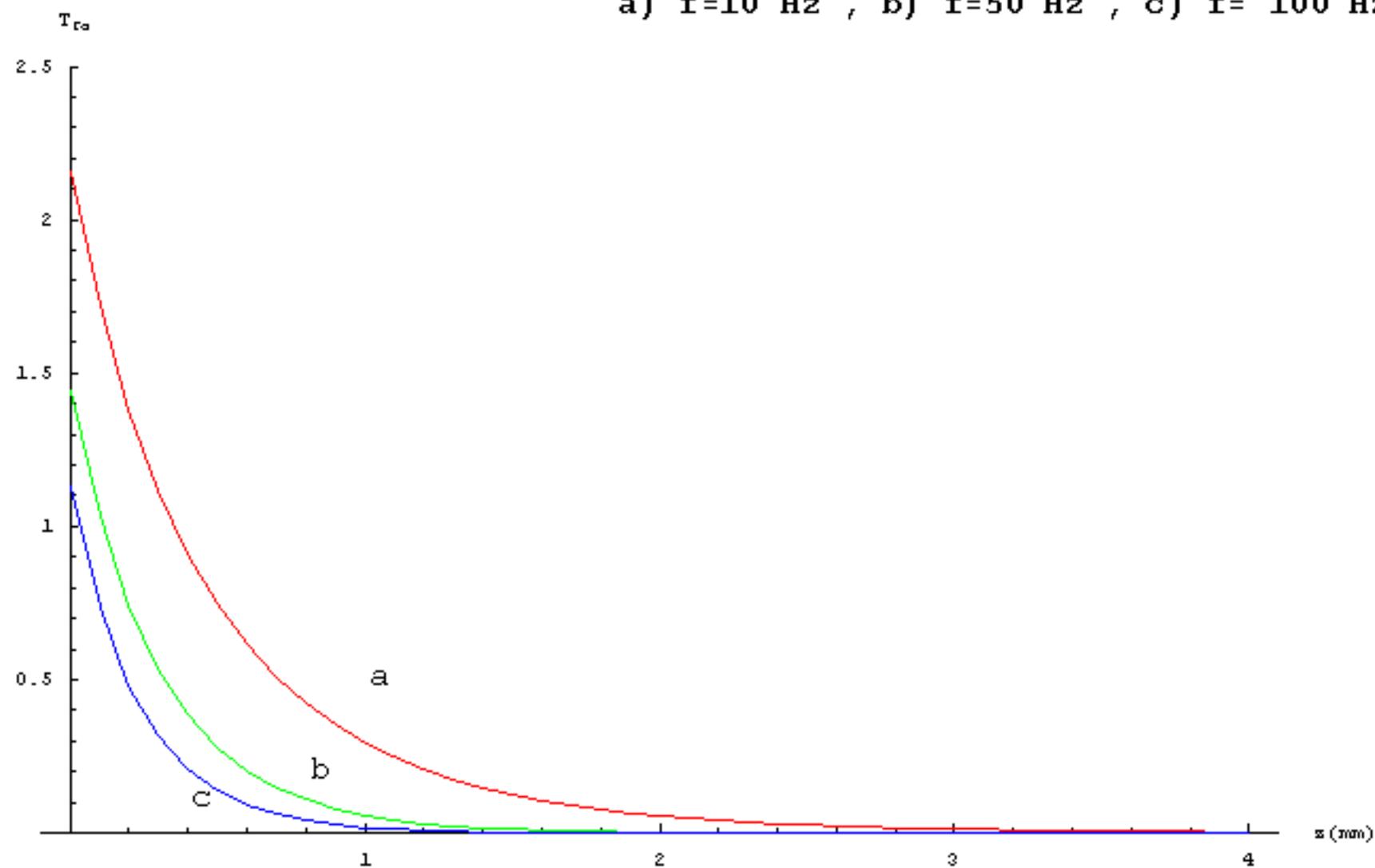

Ge : a=1 mm , l= 1 mm

a) f=10 Hz , b) f=50 Hz , c) f= 100 Hz

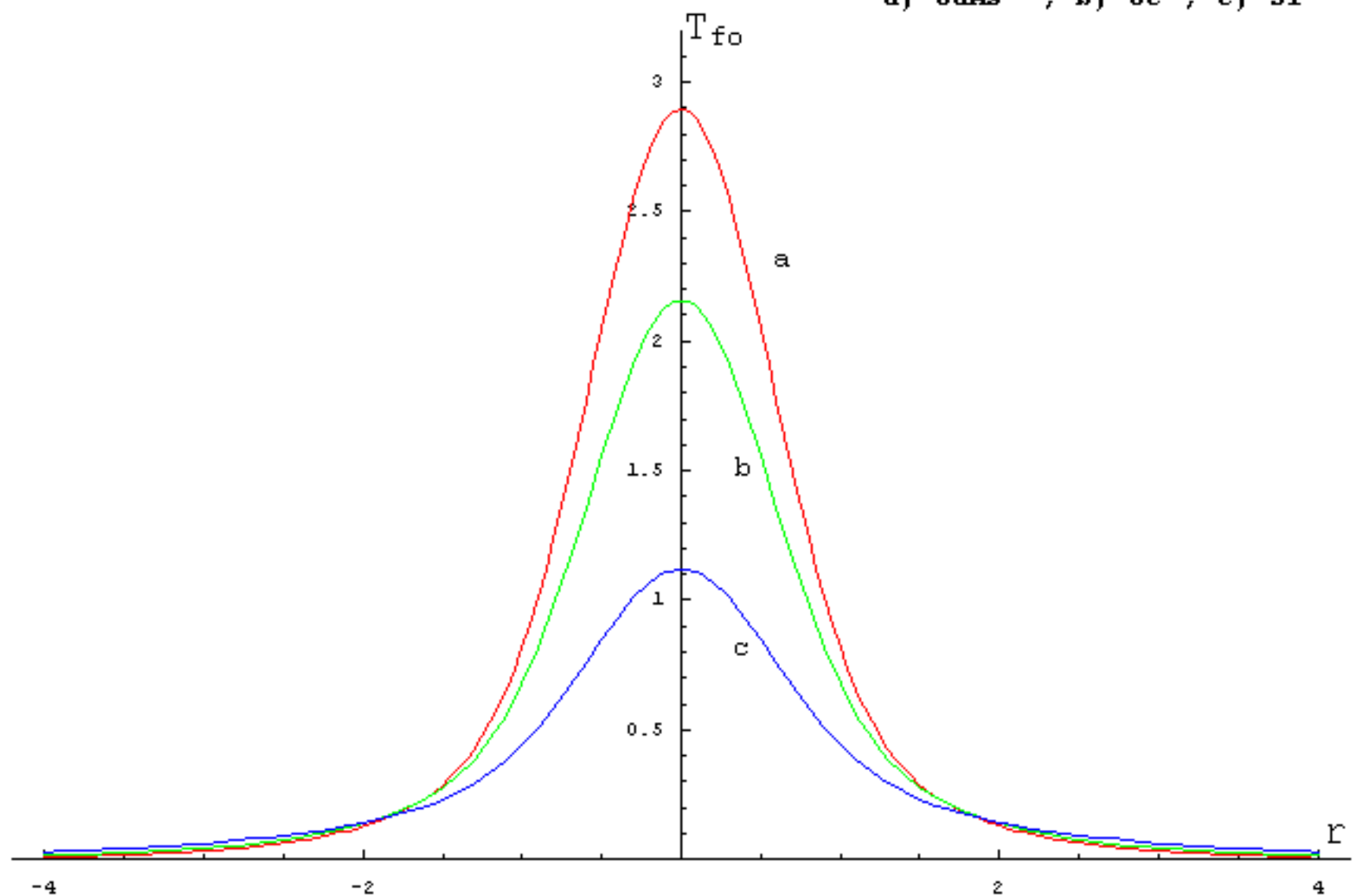

z=0 , a=1 mm , l= 1 mm , f=10 Hz

a) GaAs , b) Ge , c) Si

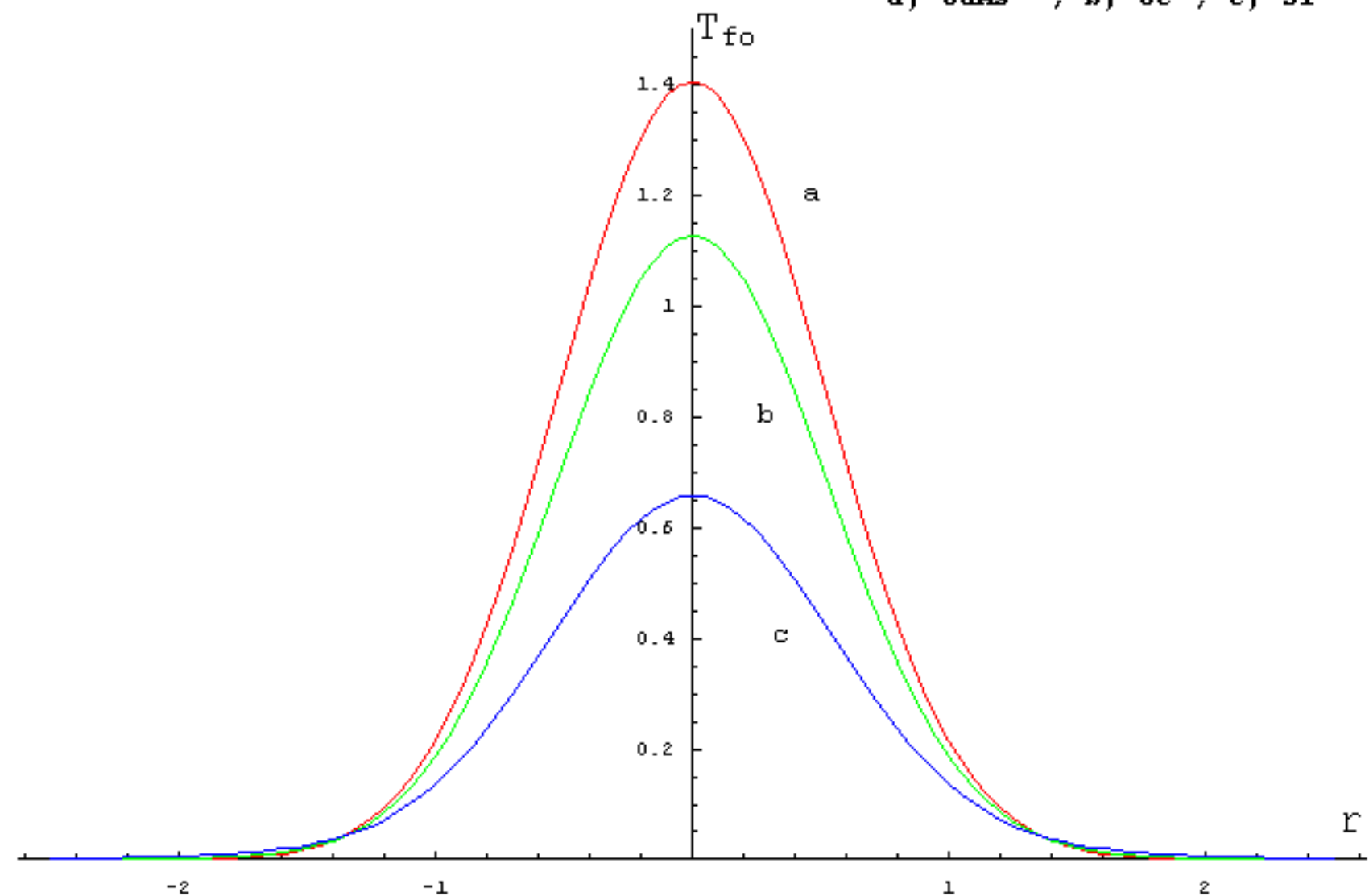

z=0 , a=1 mm , l= 1 mm , f=100 Hz

a) GaAs , b) Ge , c) Si

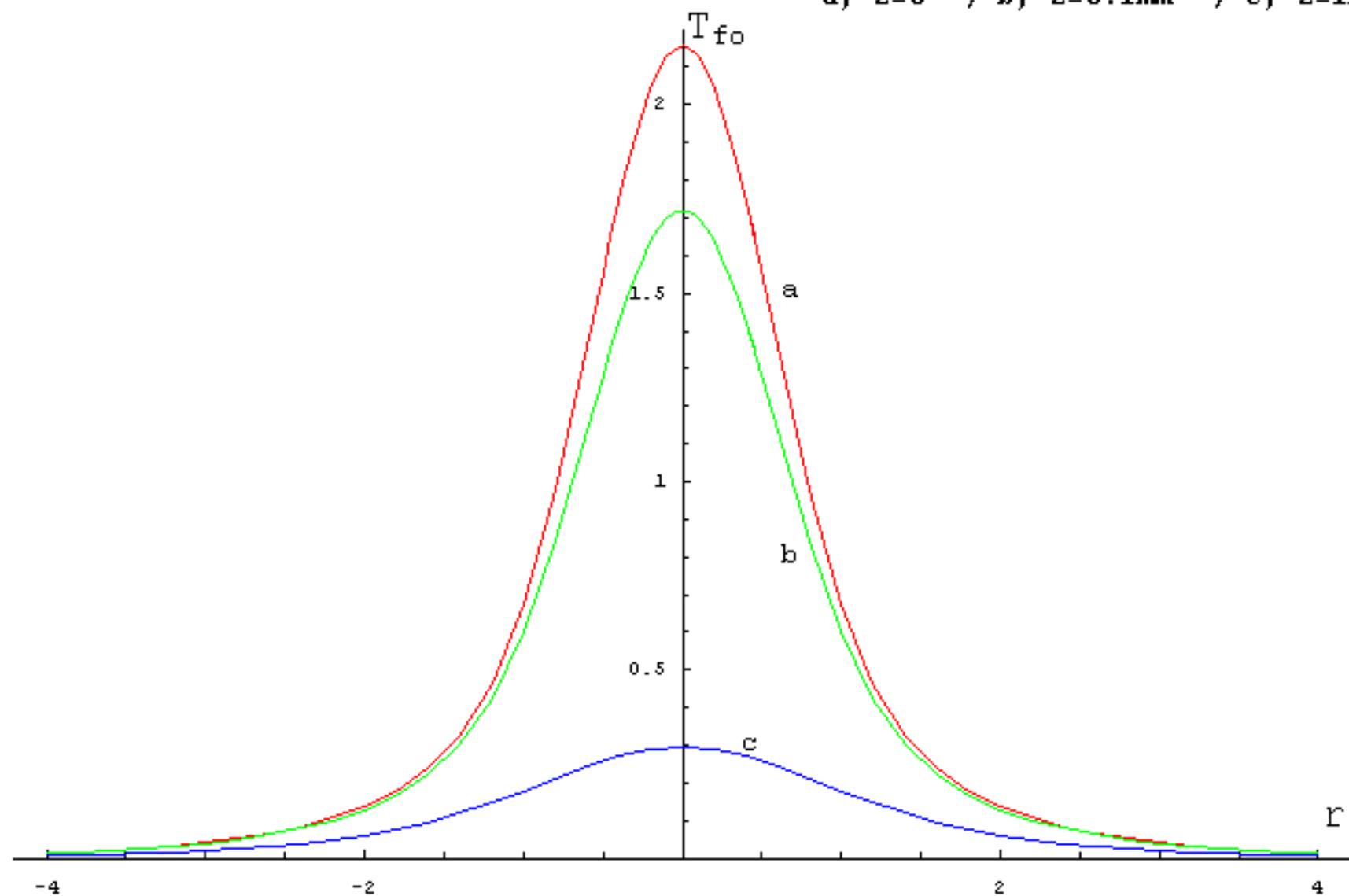

Ge : z=0 , a=1 mm , l= 1 mm , f=10 Hz

Θ_t (mR)    a) ωt=0 , b) ωt=0 +π/2 , c) ωt=0 +3π/2

0.003

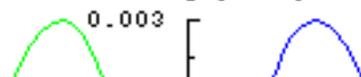

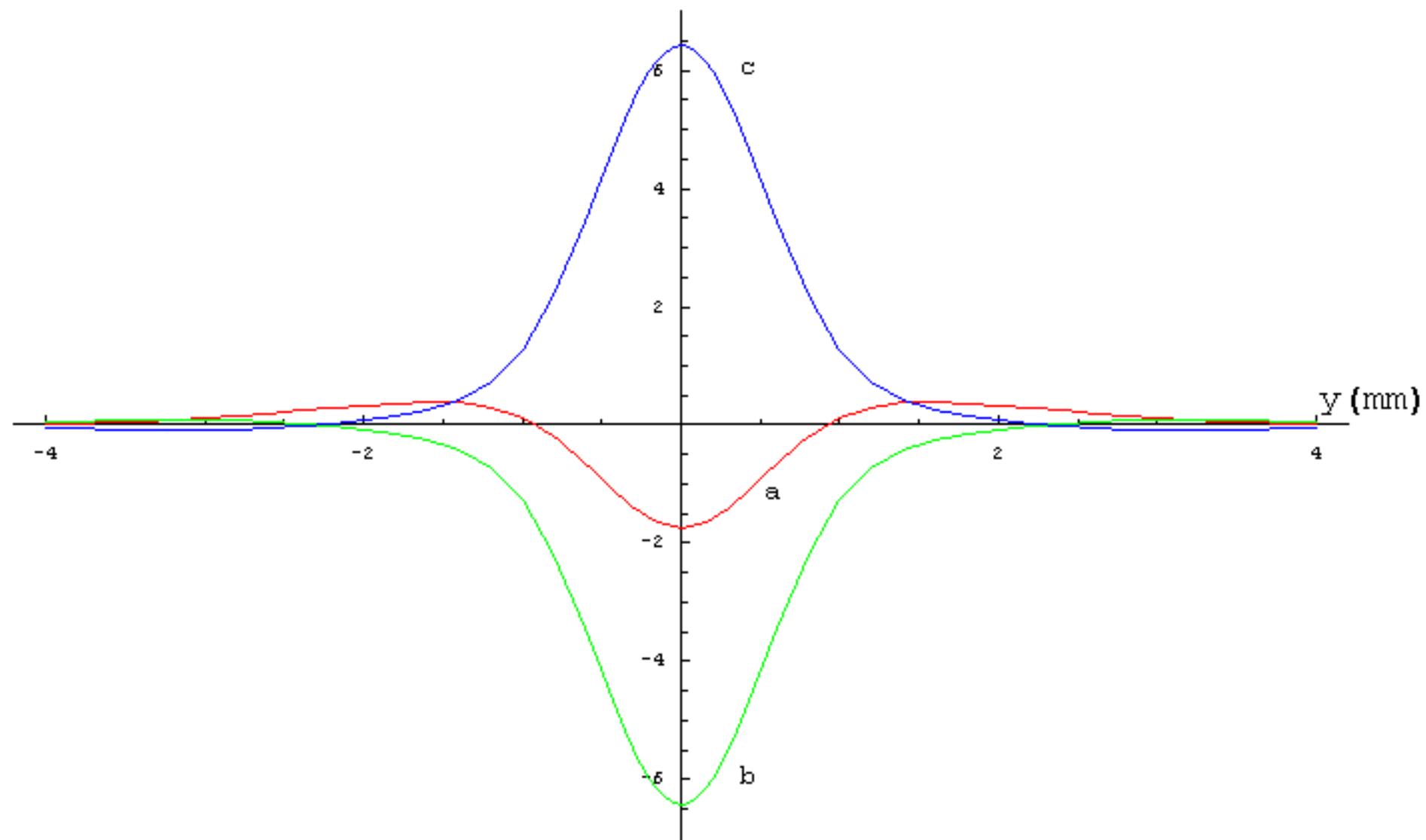

Ge : a=1 mm , l= 1 mm ,f=10 Hz

a) z=0 , b)z=0.1mm , c)z=1mm

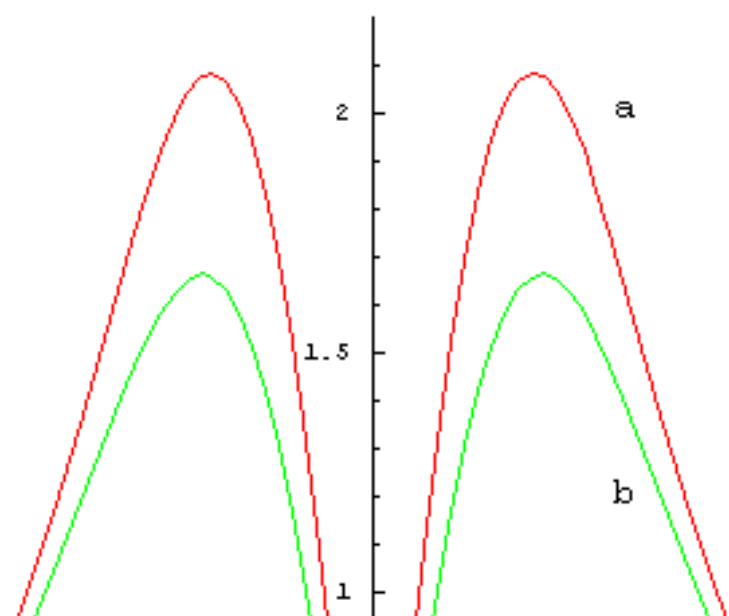

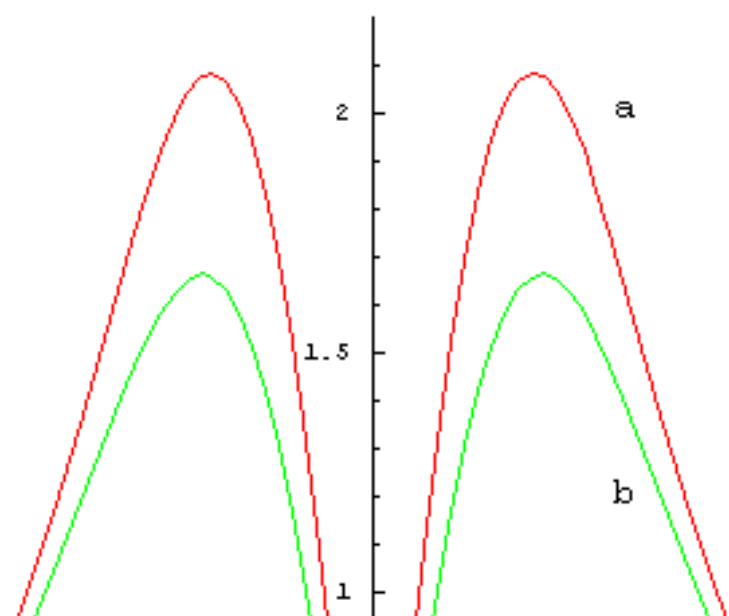

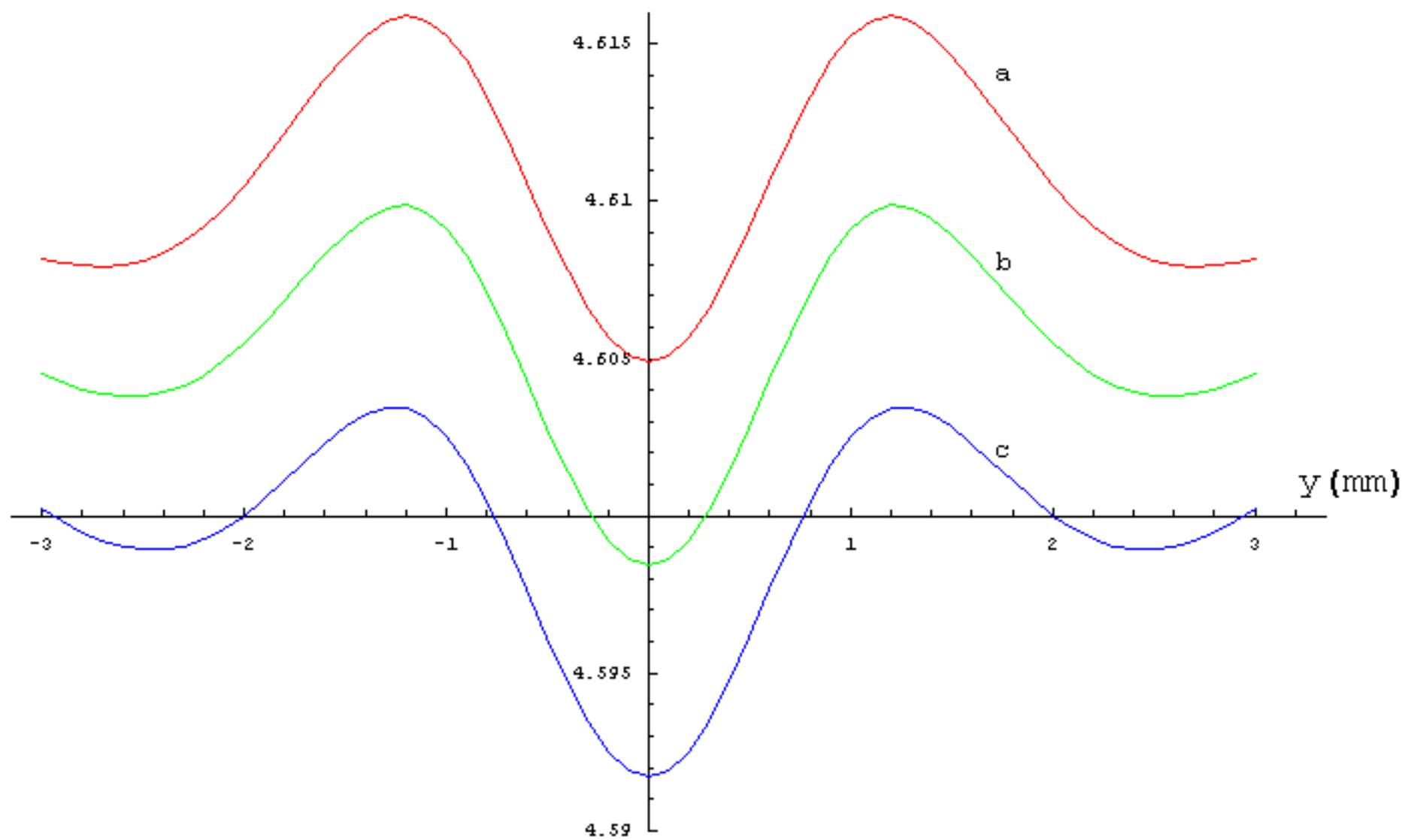

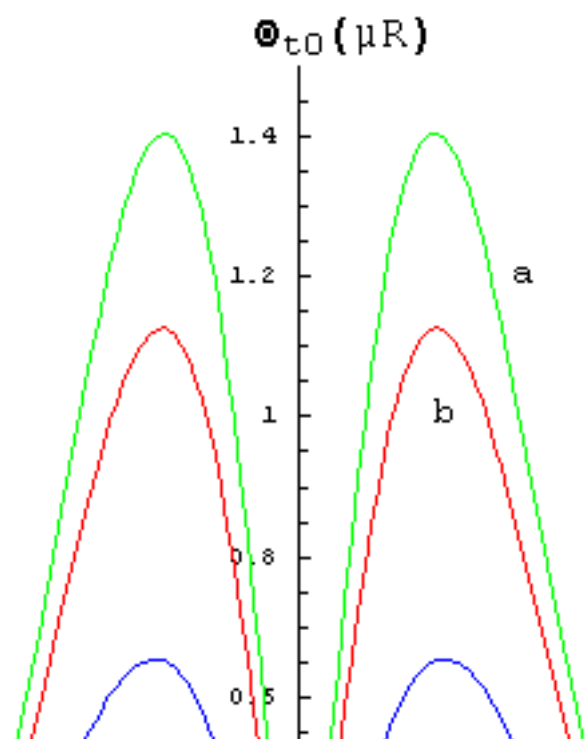

a=1 mm , l= 1 mm ,z=0 , f=100Hz

a) GaAs , b)Ge, c)Si

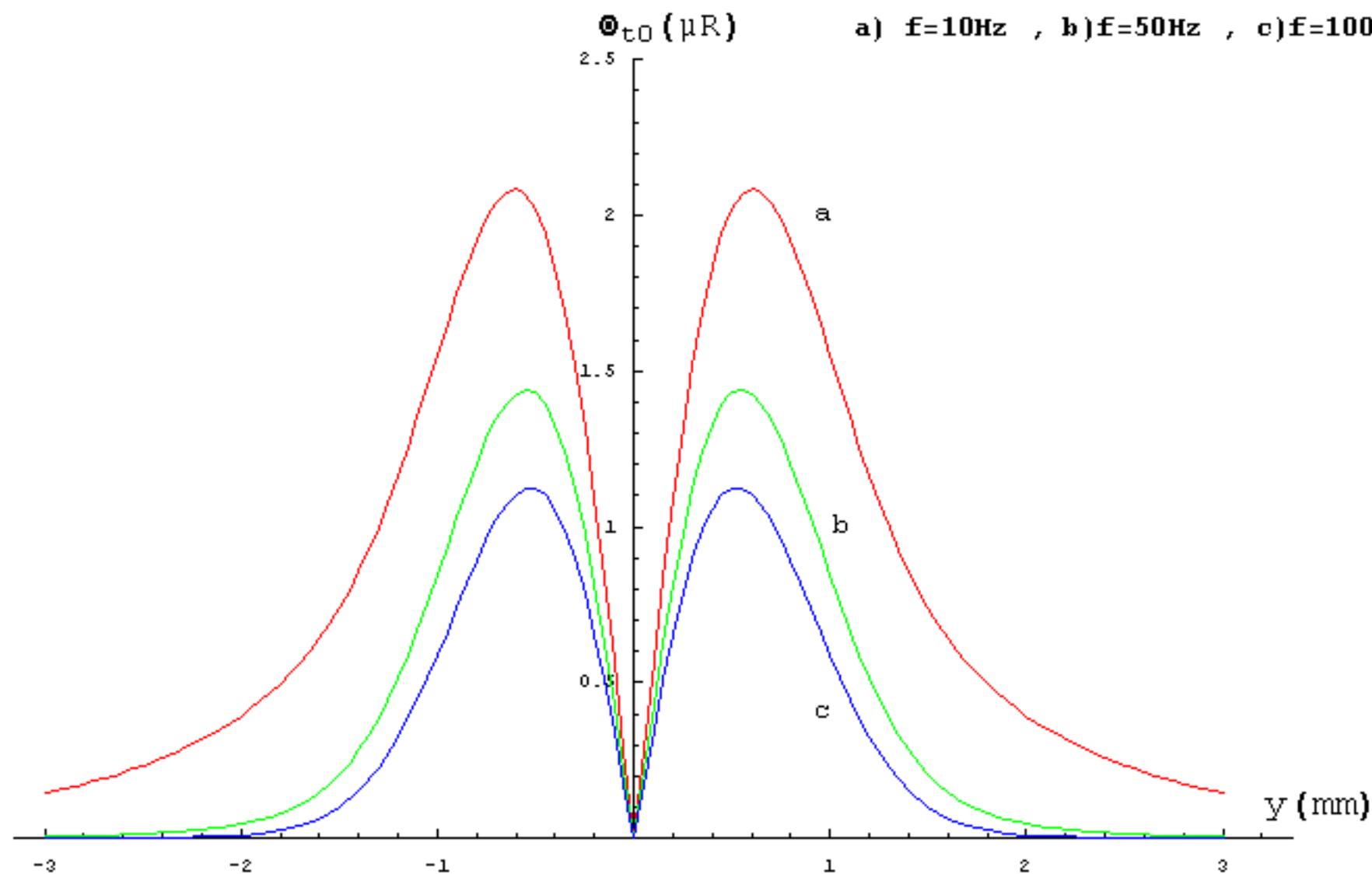